\documentclass[journal]{IEEEtran}
\usepackage{diagbox}
\usepackage{amsfonts}
\usepackage{mathrsfs}
\usepackage{cite}
\usepackage{graphicx}
\usepackage{subfigure}
\usepackage{amsmath}
\usepackage{amssymb}
\usepackage{stfloats}
\usepackage{subeqnarray}
\usepackage{cases}
\usepackage{indentfirst}
\usepackage{bm}
\usepackage{setspace}
\usepackage{url}
\usepackage{algorithmic}
\usepackage{algorithm}
\usepackage{color}
\usepackage{nomencl} 
\usepackage{parallel}
\usepackage{multirow}
\ifCLASSINFOpdf
\else
\fi
\begin{document}
%
\title{Dimensional Pricing Based on Dynamism of Load}


\author{
Jinghuan~Ma, Jie~Gu, and Zhijian~Jin

\thanks{
J.~Ma, J.~Gu and Z.~Jin are with the School of Electronic Information and Electrical Engineering,
Shanghai~Jiao~Tong~University, Shanghai, China, 200240 (email: mjhdtc@sjtu.edu.cn).
}
}


%


\maketitle

\begin{abstract}
Electricity supply is not simply a matter of quantity, but a time lasting service that matches with a wave-like load curve.
It logically deserves a pricing based on the curve per se rather than simply integral of load.
This paper introduces to price electricity consumption based on dynamism of load, which is an equivalent characterization of load.
Orthonormal basis with definable and distinguishable periodicity that can linearly express load curves constitutes a space to capture the dynamism,
where coefficients of the basis quantify the dynamism in multiple dimensions. A price function is proposed to map the coefficients to a numerical charge.
A pricing model on the space specialized by the Fourier series is derived for the simplest one-source-one-subscriber system and generalized to a single-bus system with multiple sources and multiple subscribers. Examples will demonstrate the use of the proposed pricing
and its effectiveness in reflecting the cost of generation to cope with load dynamism and guaranteeing fairness.

\begin{IEEEkeywords}
Power system economics, electricity pricing, space of load, space of load dynamism, functional analysis.
\end{IEEEkeywords}

\end{abstract}


%
\IEEEpeerreviewmaketitle

\section{Introduction}

Uniqueness of electric energy as a product is seen from both its instantaneous generation by conversion of energy of other forms and the delivery and consumption occurring simultaneously with the generation~\cite{PG-2013}. It has posed a demanding requirement, i.e. power-on-demand operation, to electric power system, which is achieved by a complex generation subsystem composed of generators with classified capacities, response rates and cost characteristics~\cite{PG-2013}.

Because of power on demand, the cost to generate electricity is not only determined by factors associated with the generation itself, e.g. cost of fuel and labor, but also by the demand side, specifically the dynamism of load and the amount of electricity consumed, which is distinguished from any other production.
But the uniqueness seems to have little inspiration on electricity pricing~\cite{PSE-2018}.
We have been pricing electricity at per unit of energy or power and charge the consumer based on the amount consumed or capacity provided, for it is a common sense or more of an intuitive way. Can it distinguish and convey the cost due to load dynamism?

\begin{figure}[!t]
\centering
\includegraphics[width=\linewidth]{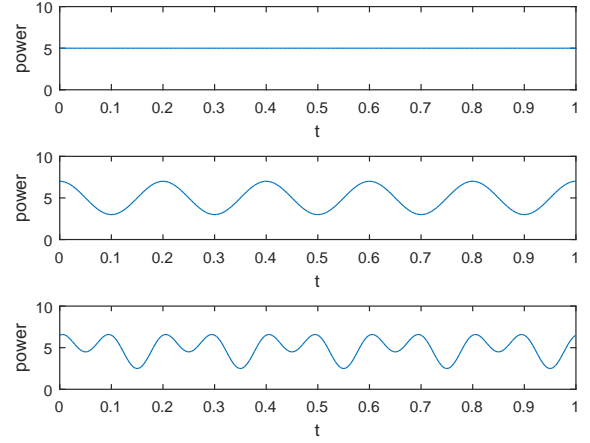}
\caption{Load curves of equal accumulated energy but different dynamism.} \label{lcuvs}
\end{figure}

In a toy example presented in Fig.~\ref{lcuvs}, three load curves\footnote{Units of measurement are reduced for convenience.} are respectively defined by $p_1(t)=5$, $p_2(t)=5+2\cos(10\pi t)$ and $p_3(t)=5+\sin(10\pi t)+1.5\cos(20\pi t)$. They consume an equal energy of 5 over $[0,1]$, but are of different dynamism.
Their payments should be equal according to the classical pricing that uses fixed unit price~\cite{EP-1982}.
However, the costs of supplier to match with the load curves cannot be equal. Generally, stable operation is more economic~\cite{PG-2013}. This means $p_2(t)$ and $p_3(t)$ probably result in higher costs. In order to prevent the supplier from suffering a loss, one can raise the unit price according to the overall cost. But the three curves still cannot be charged fairly with regard to the burden posed on system. More importantly, consumer has hardly any incentive to achieve a more system-friendly consumption pattern, i.e., one with lower dynamism. Because quantity-based charge can only adjust quantity, not the shape of load.
The factor that higher dynamism results in extra effort and cost of electricity supply has not been conveyed to the demand side.

It is worth thinking over what can be drawn from the uniqueness. To provide electric energy is not simply a matter of quantity. It is rather a time lasting service that aims to match with the demand curve of electric energy consumer.

Mathematically, load curve and generation curve belong to a square integrable space, also know as the $L_2$ space~\cite{EO-1989}.
Generation cost is indeed a functional, i.e. a function of the curve.
For convenience and application, the cost is generally approximated in two major ways. One is to count how much fuel is consumed to generate a certain amount of energy over a fixed period of time, which equals setting cost a function of the integral of power.
The other is to evaluate cost based on generation curve and averaged fuel consumption versus output power curve~\cite{PG-2013}.
In either way, the simplified cost is an average value over the corresponding time interval that reduces the complexity of the true cost as it equals using a number space to reduce the dimension of load space.

As a side effect, the cost loses information that reflects the connection between generation cost and the dynamic characteristic of generation curve.
Classical pricing that sets unit price based on the cost will never be able to effectively and fairly charge a consumer with regard to the real burden posed on system.
Because the pricing basis, integral of load, lacks too much information w.r.t. shape of load, the incentive created will never logically be capable of reshaping load.
In other words, countermeasures based on information of a multi-dimensional variable that only covers restricted dimensions is incapable to manipulate the variable in all dimensions.
Pricing strategies such as real-time pricing~\cite{DS-2011}, time-of-use pricing~\cite{TO-2012} and critical-peak pricing~\cite{RI-2007} make little difference as they just conduct dimension reduction on shorter time intervals. Besides, consumers with stable load curves will not be charged fairly as they are not the causes of cost resulting from generation fluctuation but have to pay for what they have not done.

It has been a reasonable substitution to cut a time lasting process of an object into sections and establish averaged models for the sections, or to simplify problems via dimension reduction to approximate the process, given limited ability to monitor and manage. But now there is powerful ability enabled by Internet of things technologies~\cite{IOT-2015} to monitor, communicate, compute and manage~\cite{RI-2018}. It is imperative to characterize the process of an object per se to preserve and exploit the information it contains yielding physical insights.

For power system, it is in demand to clarify the connection between generation cost and load dynamism and convey it to the demand side via certain pricing mechanism.
It thus completes the logic that a pricing based on load information covering full dimensions is logically and theoretically capable to reshape the load.

In this paper, a load curve is introduced as a square integrable function that belongs to an $L_2$ Hilbert space, named space of load.
In this space, dynamism of load can be easily and directly perceived via visual characterization as in Fig.~\ref{lcuvs}, but is hard to measure.
Instead, we propose to characterize the dynamism by a dual equivalent of the load space that is constituted by a series of orthonormal basis functions with definable and distinguishable periodicity that can linearly express load curves. We simply name the dual equivalent space of load dynamism.
Coefficients of the basis constitute a function to characterize the dynamism, whereby amplitudes and polarities of the coefficients quantify the dynamism.
Quantity, as the classical pricing basis, is actually a special case expressed by the dimension of zero dynamism.
A price functional is defined on the space of dynamism as a function of the coefficients to fully characterize the connection between generation cost and load dynamism.
Payment of electricity consumption is further determined based on the price functional to convey to the demand side the overall cost of supplier(s) to match the fluctuant demand load.

We apply the Fourier series to specialize the space of load dynamism and further quantify load dynamism.
We derive the specialized price-dynamism function and payment function for the simplest scenario, i.e., single source serving single subscriber.
To apply the pricing to a networked system, we further derive the pricing model for a single-bus system with multiple sources and multiple subscribers.

Examples demonstrates the use of the proposed pricing and its effectiveness in conveying the cost of matching load fluctuation, guaranteeing fairness of billing among subscribers and fairness of income distribution among sources.

The remainder of this paper is organized as follows.
General model is presented in Section~\ref{CGM}.
Pricing model based on the Fourier series is introduced in Section~\ref{PMFS}.
Examples are presented and analyzed in Section~\ref{EXMP}.
Conclusion is drawn in Section~\ref{clnc}.

\section{General Model}\label{CGM}
In this section, space of load is presented as an $L_2$ Hilbert space, where the elements can be decomposed into a series of orthonormal basis functions.

A dual equivalent of space of load which is constituted by orthonormal basis with definable and distinguishable periodicity is introduced to characterize fully the dynamism of load, and is simply named the space of load dynamism.

To implement the idea of charging electricity consumption based on not only the amount of energy consumed but also the dynamism of load, a general price-dynamism model is proposed for the load curve of an idea single-source-single-subscriber system, based on which a more applicable pricing scheme can be further deduced.

\subsection{Space of Load}
Let $E$ denote the energy accumulated on a time span $\Delta t$.
Classically, the rate of energy consumption, also known as power, is defined as the derivative of energy consumed with respect to time:
$dE \over dt$.
However, in this work, we define a kind of average power as
\begin{equation}
p(t,{\Delta}t)={E\big|_{ [t-{{\Delta}t\over 2},t+{{\Delta}t\over 2}] } \over {\Delta}t},
\end{equation}
where ${\Delta}t$ can be flexibly determined to fit the dynamism of the intended demand-supply scenario, i.e.,
to characterize dynamism of the object without considering overly high stochastic volatility.
We simply call $p$ \emph{power} hereinafter.

As electricity consumption can be investigated on continuous time span of any length,
time domain of the model is characterized by an arbitrary time interval $\mathcal{T}=[t_1,t_2]$ where $t_1<t_2$ holds permanently.
A load curve $$p(t), t\in\mathcal{T}$$
characterizes the entire process of an electricity consumption over $\mathcal{T}$.
The entire set of load curves is characterized by a function space
\begin{equation}
\mathcal{L}=\Big\{p(t),t\in\mathcal{T}\Big\},
\end{equation}
named the space of load.
The load curves\footnote{Properties of the load curves are introduced by \emph{Properties A.1-A.3} in Appendix~\ref{AP1}.} actually belongs to a linear function space of all real-valued functions square integrable in the interval $[t_1, t_2]$~\cite{EO-1989}, i.e.,
\begin{equation}\label{set}
\mathcal{L}\in L_2[t_1,t_2].
\end{equation}
In the space $\mathcal{L}$, we can further apply abstract definitions in $L_2[t_1,t_2]$ to reveal its good properties~\cite{EO-1989}.

\emph{Definition 1:} For $L_2[t_1,t_2]$, the inner product of $p_1(t)\in L_2[t_1,t_2]$ and $p_2(t) \in L_2[t_1,t_2]$ is defined as
\begin{equation}
\Big(p_1(t),p_2(t)\Big)=\int_{t_1}^{t_2}p_1(t)p_2(t)dt.
\end{equation}

\emph{Definition 2:} The \emph{norm} of a load curve $p(t)$ is defined as:
\begin{equation}
||p(t)||=\sqrt{\Big(p(t),p(t)\Big)}=\sqrt{\int_{t_1}^{t_2}p^2(t)dt}.
\end{equation}


\emph{Property 1:} $L_2[t_1,t_2]$ is a Hilbert space, i.e., an inner product space that is also a complete metric space w.r.t. the distance induced by the inner product.~\footnote{
For a comprehensive understanding, in Appendix~\ref{AP1}, the operation of inner product is introduced by \emph{Property A.4}; the distance induced by the inner product is introduced by \emph{Definition A.1}; and the properties of a Hilbert space are introduced by \emph{Property A.5}.}

\emph{Definition 3:} An orthonormal basis is an infinite set of functions that satisfy:
for $\varphi_i$ and $\varphi_j$ of the set,
\begin{equation}
\Big(\varphi_i,\varphi_j\Big) = \left\{ \begin{array}{ll}
1   & \text{if } i = j,\\
0  & \text{else}.
\end{array} \right.
\end{equation}

The space $L_2[t_1,t_2]$ contains many orthonormal bases such that:

\emph{Property 2:} A load curve $p(t)$ in $\mathcal{L}$ can be expressed as a linear combination of the functions of an orthonormal basis $\{ \varphi_k \}$ as:
\begin{equation}\label{series}
p(t)=\sum_k c_k\varphi_k,
\end{equation}
where
\begin{equation}
c_k=\Big(p(t),\varphi_k\Big).
\end{equation}

\emph{Property 3:} In equation~(\ref{series}), $p(t)$ and $\{c_k\}$ always satisfy the Parseval's theorem:
\begin{equation}
||p(t)||^2=\sum_k c_k^2.
\end{equation}

\subsection{Space of Load Dynamism}
Since load curve reveals the rhythm of human activities, we are interested in orthonormal bases whose basis functions have rhythmic dynamism, i.e., being periodic.
As the basis functions are normalized, their dynamism can be differentiated by the periodicity.

\subsubsection{The Orthonomal Bases}
Let $T_0=t_2-t_1$ denote length of the period. Let $f_0= {1\over T_0}$, which is referred to as the fundamental frequency over interval $\mathcal{T}$. We intend to use an orthonormal basis $\{ \phi_k|k=0,1,2,... \}$ that satisfy the following:

\emph{Property 4:}
\begin{enumerate}
\item $\phi_0$ is named the zero-frequency basis function, which satisfies
\begin{equation}
\phi_0(t)\equiv {1\over T_0}.
\end{equation}
Obviously, $\phi_0(t)$ has no dynamism and
\begin{equation}
c_0=\int_{t_1}^{t_2} p(t)dt.
\end{equation}
That is, $\phi_0$ can independently characterize energy consumption of $p(t)$ with respect to accumulation of energy consumed.
\item $\phi_k, k > 0$ satisfy
\begin{equation}
\phi_k(t+{T_0\over k})=\phi_k(t).
\end{equation}
We say that $\phi_k$ has dynamism as it fluctuates with frequency $kf_0$. We further require that
\begin{equation}
\int_{t_1}^{t_2} \phi_k(t)dt=0.
\end{equation}
That is, the basis functions with dynamism accumulate zero energy in time interval $\mathcal{T}$. Thus, dynamism is independent of energy accumulation that is solely characterized by $\phi_0$.
\end{enumerate}

For $\phi_i$ satisfying $\phi_i(t+{T_0\over i})=\phi_i(t)$ and $\phi_j$ satisfying $\phi_j(t+{T_0\over j})=\phi_j(t)$, if
\begin{equation}
i>j,
\end{equation}
we say that the dynamism of $\phi_i$ with regard to frequency is higher than that of $\phi_j$.

\subsubsection{Space of Load Dynamism}
According to equation~(\ref{series}), a load curve can be decomposed by $\{ \phi_k|k=0,1,2,... \}$ into two major parts: the accumulative part measured by $T_0 c_0 \phi_0$, the dynamic part characterized by the other basis functions $\{ \phi_k|k=1,2,... \}$. Moreover, the dynamism of a load curve has to be analyzed from more than just the aspect of frequency.
The other aspect is strength, i.e., coefficients of the basis $\{c_k,k\in \mathbb{Z}\}$ indicate the polarity and amplitude of frequency component $\phi_k$.

$\{c_k,k\in \mathbb{Z}\}$ constitute a vector $\bf c \in \mathbb{R}^n$ of multiple dimensions.
The space of the vectors is denoted by $\mathcal{H}$.
A one-to-one mapping $\mathcal{F}$ can be defined between $\bf c$ and $p(t)$ as
\begin{equation}
{\bf c}=\mathcal{F}\left(p(t)\right): \mathcal{L} \to \mathcal{H}.
\end{equation}
$\mathcal{H}$ is a dual equivalent of $\mathcal{L}$.
As the vector can fully characterize the dynamism of load, it is named the vector of load dynamism. $\mathcal{H}$ is therefore named the space of load dynamism.

\subsection{Price-Dynamism Function and Payment}
By decomposing load curve into a series of orthogonal basis functions $\{c_k\varphi_k, k\in \mathbb{Z}\}$ that are of different frequencies, polarities and amplitudes, we point out that classical pricing charges consumer mainly, probably solely, based on $\int_{t_1}^{t_2}c_0\phi_0 dt$, i.e. the amount of energy consumed.
But the effort of power supplier to cope with the dynamism of load curve, as suggested by the non-zero frequency basis functions $\{c_k\varphi_k, k\in \mathbb{Z}_+\}$, has not been, at least properly, addressed, since existing amount-based pricing~\cite{DS-2011,TO-2012,RI-2007} shifts all cost roughly into a unit price or time-of-use unit price.

\begin{figure}[!t]
\centering
\includegraphics[width=3in]{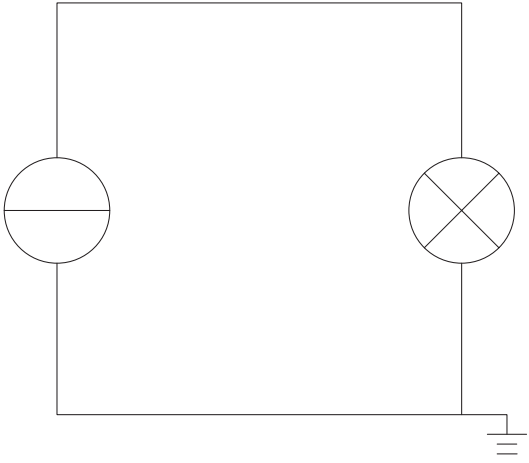}
\caption{A simplest source-load pair.} \label{model}
\end{figure}

To convey the reality and create a strong incentive for the demand side to reduce consumption dynamism appeared to the supplier, we propose a general price-dynamism function for an ideal model in Fig.~\ref{model} that is the simplest source-load pair.
The source is assumed to always match the load at any cost.
As the dynamism is captured by multiple dimensions, the price is therefore a combination of multiple dimensions.
Each of the orthonormal basis $\{ \phi_k|k=0,1,2,... \}$ indicates the dynamism of $\phi_k$ with unit amplitude.
A unit price denoted by $\lambda$, named price-dynamism coefficient, is defined for each of the basis functions as
\begin{equation}
\lambda_k=\lambda({kf_0}),
\end{equation}
where the mapping $\lambda(\cdot)$ is named price-dynamism function. The price-dynamism function can be applied to both the supply side and the demand side, separately, to bear different meanings.
\begin{enumerate}
\item At the supply side, $\lambda(\cdot)$ can be used equally as cost-dynamism function by simply setting $\lambda_k$ the cost to cope with dynamism of $\phi_k$ of unit strength. $\lambda(\cdot)$ indicates the dynamic characteristic of a source, which may be approximated via experiment and observation.
\item At the demand side, $\lambda(\cdot)$ can be used to determine the unit price $\lambda_k$ that a consumer pays for its load dynamism that equals $\phi_k$ with unit strength.
Here, $\lambda(\cdot)$ can be set in accordance with the dynamic characteristics of related sources. The total payment of a load curve, denoted by $P$, is defined as:
\begin{equation}\label{pay}
P= \lambda_0 \int_{t_1}^{t_2}c_0\phi_0 dt+ T_0\sum_{k,k>0} \lambda_k c_k.
\end{equation}
In equation~(\ref{pay}), $\lambda_0$ and $\int_{t_1}^{t_2}c_0\phi_0 dt$ can be regarded as the classical unit price and the total payment for the amount of energy consumed, respectively. $\lambda_k,k>0$ is the unit payment for adding dynamism to the system captured by $c_k\phi_k,k>0$.
This suggests that the proposed dimensional pricing can be regarded as a generalization of the classical pricing.
\end{enumerate}

\section{Pricing Model based on the Fourier Series}\label{PMFS}
The Fourier series~\cite{EO-1989} has been widely applied to analyze electrical phenomena. Moreover, it can distinguish dynamism of a load curve from the energy accumulated as the Fourier series satisfy \emph{Property 4}.

In this section, we first introduce the Fourier Series as a special case of $\{ \phi_k|k=0,1,... \}$.
Based on them we further present a pricing model not only for the simplest source-load pair, but also for a single-bus ideal system with multiple sources and multiple subscribers.

\subsection{The Fourier Series}
In $L_2[t_1,t_2]$, a set of functions
\begin{small}
\begin{equation}\nonumber
{1 \over T_0}, \cos {2\pi\over T_0} t, \sin {2\pi \over T_0} t,\cos {4\pi \over T_0} t, \sin {4\pi \over T_0} t,...\cos 2\pi{n \over T_0} t, \sin 2\pi {n\over T_0} t,...
\end{equation}
\end{small}
constitute an orthonormal basis. Each of the basis functions is also called a frequency component. ${1 \over T_0}$ is the zero-frequency component. $\cos {2\pi\over T_0} t$ and $\sin {2\pi \over T_0} t$ are the fundamental frequency components. The other components each has an integer multiples of the fundamental frequency.

A load curve $p(t)\in \mathcal{L}$, as it naturally satisfies the Dirichlet conditions~\footnote{The Dirichlet conditions is the sufficient conditions for a real-valued function on an intended interval to be equal to the sum of its Fourier series. Please refer to \emph{Property A.6} in Appendix~\ref{AP1}.}, can be written as:
\begin{equation}\label{FS}
p(t)={a_0\over 2}+\sum_{n=1}^\infty\Big[a_n\cos(2\pi{n f_0}t)+b_n\sin(2\pi{n f_0}t) \Big],
\end{equation}
where $n\in \mathbb{Z}_+$, coefficients
\begin{equation}
a_n={2\over T_0}\int_{t_1}^{t_2} p(t)\cos(2\pi{n f_0}t)dt
\end{equation}
and
\begin{equation}
b_n={2\over T_0}\int_{t_1}^{t_2} p(t)\sin(2\pi{n f_0} t)dt
\end{equation}
respectively indicate the polarities and amplitudes of the corresponding sine and cosine components.
The right side of equation~(\ref{FS}) is called the Fourier series. Obviously,
\begin{equation}
a_0 ={2\over T_0}\int_{t_1}^{t_2} p(t)dt,~\footnote{Generally, we use $1\over 2$ instead of $1\over T_0$ as the zero-frequency basis function in a Fourier series.}
\end{equation}
and ${T_0\over 2}a_0$ equals the energy consumed. Hence, the Fourier coefficients $\{(a_n,b_n)|n\in\mathbb{Z}\}$ fully characterize a load curve w.r.t. energy consumed and the dynamism.

\subsection{The Pricing}

\subsubsection{Price-Frequency Coefficients}
For the zero-frequency component, let $\alpha_0 >0$ denote the price per unit of energy consumed. For the cosine components, let $\alpha_n, n>0$ represent the unit price for $a_n\cos(2\pi{n f_0}t)$. For the sine components, let $\beta_n, n>0$ represent the unit price for $b_n\sin(2\pi{n f_0}t)$.

For the simplest source-pair case presented in Fig.~\ref{model}, it is easy to calculate the payment of a subscriber according to
\begin{equation}\label{pay2}
P= \alpha_0 {T_0\over 2}a_0 + T_0\sum_{n=1}^{\infty} (\alpha_n a_n+\beta_n b_n).
\end{equation}

\begin{figure}[!t]
\centering
\includegraphics[width=\linewidth]{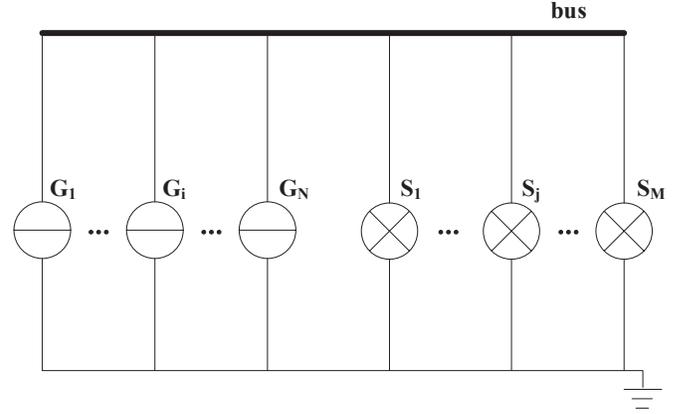}
\caption{A single-bus ideal system with multiple generators and multiple subscribers.} \label{bus}
\end{figure}

We further consider a more complex case as presented in Fig.~\ref{bus}, i.e., a single-bus ideal system with multiple generators and multiple subscribers. The term ``ideal'' refers to that in the system the length of power line is ignored, let alone the power loss and the change of phase. We recall that the load curves of the participants are independent. Hence, the analysis is tripartite: one source serving multiple subscribers, one subscriber served by multiple sources and multiple sources serving multiple subscribers.

\subsubsection{One-Source-Multiple-Subscriber}
We assume that only source $i$ generates electricity while the other sources do not.
For a source $i$, its generation curve $g_i(t)$ equals sum of the partial load curves of which the subscribers are served by $i$:
\begin{equation}\label{SGi}
g_i(t)=\sum_{j=1}^M p_{ji}(t),
\end{equation}
where $p_{ji}(t)$ denotes the partial load curve of subscriber $j$ served by source $i$.
For each of the subscribers, $p_{ji}(t)$ can be expressed by the Fourier series as:
\begin{equation}\label{FSji}
p_{ji}(t)={a_{j0i}\over 2}+\sum_{n=1}^\infty\Big[a_{jni}\cos(2\pi{n f_0}t)+b_{jni}\sin(2\pi{n f_0}t) \Big],
\end{equation}
where the subscript triple $\cdot_{jni}$ refers to subscriber $j$, frequency index $n$ and source $i$.
Hence, we can rewrite $g_i(t)$ as
\begin{equation}\label{SGidecomp}
\begin{aligned}
&g_i(t)=\\
&\sum_{j=1}^M \Bigg[ {a_{j0i}\over 2}+\sum_{n=1}^\infty \Big[a_{jni}\cos(2\pi{n f_0}t)+b_{jni}\sin(2\pi{n f_0}t) \Big]\Bigg]\\
&={\sum_{j=1}^M a_{j0i}\over 2}+\\
&\;\;\;\;\;\sum_{n=1}^\infty \Big[ (\sum_{j=1}^M a_{jni})\cos(2\pi{n f_0}t)+(\sum_{j=1}^M b_{jni})\sin(2\pi{n f_0}t) \Big].
\end{aligned}
\end{equation}
A set of price-dynamism coefficients are denoted by:
\begin{equation}\label{pricei}
\Big\{ \alpha_{i0}, \{\alpha_{in}|n\in \mathbb{Z}_+\}, \{\beta_{in}|n\in \mathbb{Z}_+\} \Big\},
\end{equation}
which are set to reflect the dynamic characteristics of source $i$.
The total payment ${\rm P}_{i}$ at the aggregate level is thus given by:
\begin{equation}\label{payalli}
{\rm P}_{i}=\alpha_{i0} {T_0\over 2} \sum_{j=1}^M a_{j0i} + T_0\sum_{n=1}^{\infty} (\alpha_{in} \sum_{j=1}^M a_{jni}+\beta_{in} \sum_{j=1}^M b_{jni}).
\end{equation}
The payment of subscriber $j$ associated with source $i$ is calculated as
\begin{equation}\label{payji}
P_{ji}= \alpha_{i0} {T_0\over 2} a_{j0i} + T_0\sum_{n=1}^{\infty} (\alpha_{in} a_{jni}+\beta_{in} b_{jni}).
\end{equation}

\subsubsection{Multiple-Source-One-Subscriber}
We assume only subscriber $j$ consumes electricity while the others do not.
For a subscriber $j$, its load curve $p_j(t)$ can be expressed by the Fourier series as:
\begin{equation}\label{FSj}
p_j(t)={a_{j0}\over 2}+\sum_{n=1}^\infty\Big[a_{jn}\cos(2\pi{n f_0}t)+b_{jn}\sin(2\pi{n f_0}t) \Big].
\end{equation}
On the other hand, subscriber $j$ is served by $I$ independent sources as in Fig.~\ref{bus}, which can be mathematically expressed by:
\begin{equation}\label{SGj}
p_j(t)=\sum_{i=1}^I p_{ji}(t),
\end{equation}
where $p_{ji}(t)$ is the partial load curve served by source $i$.
Since the price-dynamism coefficients convey the cost of electricity supply and the sources are different w.r.t. characteristics of fuel consumption, each component $p_{ji}(t)$ should be charged based on an independent set of price-dynamism coefficients defined in~(\ref{pricei}) associated with the characteristics of source $i$.
Consequently, frequency components in equation~(\ref{FSj}) have to be further decomposed based on different sources and rewritten as:
\begin{equation}\label{FSjn}
\begin{aligned}
&p_j(t)=\\
&\sum_{i=1}^I \Bigg[ {a_{j0i}\over 2} +\sum_{n=1}^\infty\Big[a_{jni}\cos(2\pi{n f_0}t)+b_{jni}\sin(2\pi{n f_0}t) \Big]\Bigg].
\end{aligned}
\end{equation}
Therefore, the payment of subscriber $j$ is determined by
\begin{equation}\label{payj}
P_j= \sum_{i=1}^I \Big[  \alpha_{i0} {T_0\over 2} a_{j0i} + T_0\sum_{n=1}^{\infty} (\alpha_{in} a_{jni}+\beta_{in} b_{jni}) \Big].
\end{equation}

\subsubsection{Multiple-Source-Multiple-Subscriber}
In such a case, it might be hard for source $i$ to assess $p_{ji}(t)$ in equation~(\ref{SGi}), and for subscriber $j$ to assess $p_{ji}(t)$ in equation~(\ref{SGj}).
This means $\big\{ \{a_{j0i},a_{jni},b_{jni}\} |i=1,...,I;j=1,...,M \big\}$ might be unavailable. Instead, we present a method to determine the payment of subscriber $j$ based on
$\big\{ \{a_{0i},a_{ni},b_{ni}\} |i=1,...,I\big\}$ and $\big\{ \{a_{j0},a_{jn},b_{jn}\} | j=1,...,M \big\}$.
At bus level, we have:
\begin{equation}
\begin{aligned}
\sum_{i=1}^I a_{0i}=\sum_{j=1}^M a_{j0},\\
\sum_{i=1}^I a_{ni}=\sum_{j=1}^M a_{jn},\\
\sum_{i=1}^I b_{ni}=\sum_{j=1}^M b_{jn}.
\end{aligned}
\end{equation}
For pricing, we propose a series of equivalent coefficients $\alpha_0'$, $\alpha_n'$ and $\beta_n'$, satisfying
\begin{equation}\label{equi}
\begin{aligned}
\sum_{i=1}^I \alpha_{i0} a_{0i}= \alpha_{0}' \sum_{j=1}^M a_{j0},\\
\sum_{i=1}^I \alpha_{in} a_{ni}= \alpha_{n}' \sum_{j=1}^M a_{jn},\\
\sum_{i=1}^I \beta_{in} b_{ni}= \beta_{n}' \sum_{j=1}^M b_{jn}.
\end{aligned}
\end{equation}
Hence, the payment of subscriber $j$ can be calculated according to:
\begin{equation}\label{paysmp}
P_j= \alpha_0' {T_0\over 2}a_{j0} + T_0\sum_{n=1}^{\infty} (\alpha_n' a_{jn}+\beta_n' b_{jn}).
\end{equation}
Equations~(\ref{payji}), (\ref{payj}) and (\ref{paysmp}) provide a rigorous perspective to link price to the real state of power system, since at whatever level the price is subdivided into, we can always match the subdivided parts to real physical meanings.

\section{Examples}\label{EXMP}
In this section, continuous examples are presented to illustrate principles of the proposed pricing at a theoretical level, where practical issues such as units of measurement and physical bases of price-frequency functions are not considered. All load curves are defined over $[0,1]$, which means $T_0=1$.

\subsection{One-Source-One-Subscriber}\label{suba}
\begin{figure}[!t]
\centering
\includegraphics[width=\linewidth]{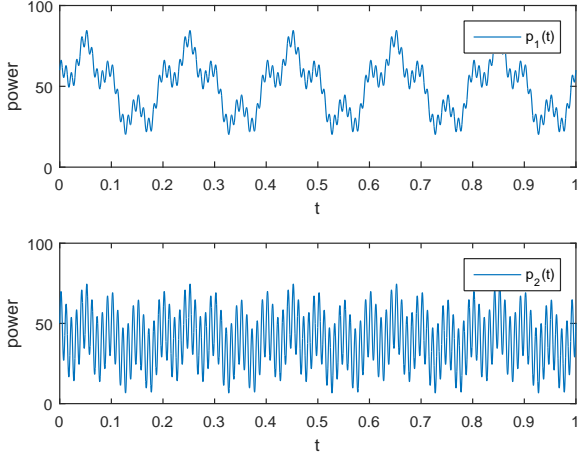}
\caption{Load prifles:$p_1(t)=50+20\sin(10\pi t)+10\cos(40\pi t)+5\sin(200\pi t)$, $
p_2(t)=40+5\sin(10\pi t)+10\cos(40\pi t)+20\sin(200\pi t)$.} \label{sload}
\end{figure}

For comparison, we assess the payments of two load curves that are respectively defined by
\begin{equation}
\begin{aligned}
p_1(t)=50+20\sin(10\pi t)+10\cos(40\pi t)+5\sin(200\pi t),\\
p_2(t)=40+5\sin(10\pi t)+10\cos(40\pi t)+20\sin(200\pi t),
\end{aligned}
\end{equation}
which are drawn in Fig.~\ref{sload}.
For $p_1(t)$, $(a_0)_1=100$, $(b_5)_1=20$, $(a_{20})_1=10$ and $(b_{100})_1=5$.
For $p_2(t)$, $(a_0)_2=80$, $(b_5)_2=5$, $(a_{20})_2=10$ and $(b_{100})_2=20$.
Obviously, Load 2 consumes less energy than Load 1 but is of higher dynamism.

\begin{figure}[!t]
\centering
\includegraphics[width=\linewidth]{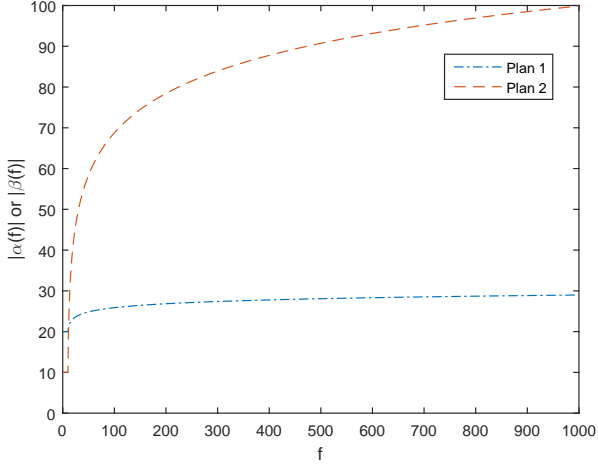}
\caption{Price-frequency coefficient functions.} \label{ss}
\end{figure}

We also assume two billing plans, i.e. two sets of price-frequency coefficient functions that reflect the dynamic characteristics of two different sources, respectively.
As shown in Fig.~\ref{ss}, we define $\{[|\alpha(f)|]_1,[|\beta(f)|]_1\}$ for Plan 1 as:
\begin{equation}
\begin{aligned}
\;\;[|\alpha(f)|]_1&=[|\beta(f)|]_1\\
&=\left\{ \begin{array}{ll}
20   & \text{if } 0\le f < 10,\\
3\log_{10}(f-9)+20  & \text{else}.
\end{array} \right.
\end{aligned}
\end{equation}
and
$\{[|\alpha(f)|]_2,[|\beta(f)|]_2\}$ for Plan 2 as:
\begin{equation}
\begin{aligned}
\;\;[|\alpha(f)|]_2&=[|\beta(f)|]_2\\
&=\left\{ \begin{array}{ll}
10   & \text{if } 0\le f < 10,\\
30\log_{10}(f-9)+10  & \text{else}.
\end{array} \right.
\end{aligned}
\end{equation}
They are presented in the form of absolute value.
To determine the final value of a price-frequency coefficient, we assume that the polarity of the price-frequency coefficient should be same with that of the corresponding Fourier coefficient at the supply side, e.g. $\alpha(f)a(f)>0$ and $\beta(f)b(f)>0$.\footnote{The assumption is made for presentation of the basic idea at a theoretical
level. We note that $\alpha(f)a(f)>0$ or $\beta(f)b(f)>0$ may not always hold in practice. But so far no experiment has been conducted to study the relationship between cost-frequency coefficients and the corresponding Fourier coefficients w.r.t. the dynamic characteristics of a generator.}
In a one-source-one-subscriber case, the generation curve equals the load curve. Hence, for Plan 1, $[\alpha(0)]_1=20$, $[\beta(5)]_1=20$, $[\alpha(20)]_1=23.9031$ and $[\beta(100)]_1=26$.
For Plan 2, $[\alpha(0)]_2=10$, $[\beta(5)]_2=10$, $[\alpha(20)]_2=49.0309$ and $[\beta(100)]_1=70$.
Compared to Plan 1, Plan 2 charges for the energy consumed at a lower rate but charges for the dynamism at a much higher rate.

Bills of the four combinations are presented in TABLE~\ref{paytbl}. Since Load 1 consumes more electricity than Load 2, Load 1 is charged more than Load 2 for the non-dynamic part by either plan. Since Load 2 poses heavier burdens on the system than Load 1 w.r.t. dynamism of load, Load 2 is charged more than Load 1 for the dynamic part by either plan.
It suggests that the source represented by Plan 1 is more cost-efficient in serving load of high dynamism, e.g., Load 2, while that represented by Plan 2 is more cost-effective in serving load of low dynamism, e.g., Load 1.

\begin{table}
\caption{Bills of the four combinations}
\label{paytbl}
\centering
\begin{tabular}{|l|c|c|c|}
\hline
\diagbox{Combo.}{Value}{Item} & Non-dyn. & Dyn.& Total \\
\hline
Load1Plan1 & 1000 & 769.031  & 1769.031\\
\hline
Load2Plan1 &  800 & 859.031 & 1659.031 \\
\hline
Load1Plan2 &  500 & 1040.309&  1540.309\\
\hline
Load2Plan2 &  400 & 1940.309 & 2340.309\\
\hline
\end{tabular}
\end{table}

\subsection{Multiple-Source-One-Subscriber}
We assume that Load 1 in Section~\ref{suba} is now served by 2 independent sources.
Source 1 charges the subscriber by Plan 1 and its generation curve $g_1(t)$ is assumed to be
\begin{equation}
\begin{aligned}
g_1(t)=15+10\cos(40\pi t)+5\sin(200\pi t).\\
\end{aligned}
\end{equation}
We have $a_{1\;0\;1}=30$, $a_{1\;20\;1}=10$ and $b_{1\;100\;1}=5$. The payment to Source 1 is $669.031$.
Source 2 charges the subscriber by Plan 2 and its generation curve $g_2(t)$ is assumed to be
\begin{equation}
\begin{aligned}
g_2(t)=35+20\sin(10\pi t).\\
\end{aligned}
\end{equation}
We have $a_{1\;0\;2}=70$ and $b_{1\;5\;2}=20$. The payment to Source 2 is $750$.
The total payment of Load 1 is $1419.031$ which is lower than Load1Plan1 and Load1Plan2, as it takes advantage of the dynamic characteristics of both sources.

\subsection{One-Source-Multiple-Subscriber}\label{subc}
We assume three independent load curves that are respectively defined by:
\begin{equation}
\begin{aligned}
p_3(t)=30+15\cos(40\pi t)+9\sin(40\pi t),\\
p_4(t)=40+15\cos(40\pi t)+5\sin(40\pi t),\\
p_5(t)=50-25\cos(40\pi t)-15\sin(40\pi t).\\
\end{aligned}
\end{equation}
Hence, the generation curve is
\begin{equation}\label{gt}
g(t)=\sum_{n=3}^5 p_n(t)=120+5\cos(40\pi t)-\sin(40\pi t).
\end{equation}
The price-frequency coefficients are determined according to the generation curve. We assume that $\alpha(0)=20$, $\alpha(20)=20$ and $\beta(20)=-25$.
Breakdown of the three payments are summarized in TABLE~\ref{paytbl1}.
We can see that the polarity of Fourier coefficients of a same trigonometric function actually indicates whether it is an aggravation to the aggregated load or a compensation.
The price-frequency coefficient of dynamic components times the corresponding Fourier coefficient functions partially as a kind of transfer payment.
It charges those who aggravates the fluctuation and reward those who compensates w.r.t. the aggregated load curve.


\begin{table}
\caption{Breakdown of payments}
\label{paytbl1}
\centering
\begin{tabular}{|l|c|c|c|}
\hline
\diagbox{Item}{Value}{Load} & Load3 & Load4 & Load5\\
\hline
$a_0$               & 60   & 80  & 100\\
\hline
$a_0$ payment       & 600  & 800 & 1000\\
\hline
$a_{20}$            & 15   & 15  & -25\\
\hline
$a_{20}$ payment    & 300  & 300 & -500\\
\hline
$b_{20}$            & 9    & 5   & -15\\
\hline
$b_{20}$ payment    & -225 &-125 & 375\\
\hline
Total               & 675  & 975 & 875\\
\hline
\end{tabular}
\end{table}

\subsection{Multiple-Source-Multiple-Subscriber}
We assume that subscribers in Section~\ref{subc} are now served by three independent sources, i.e., $g(t)$ in equation~(\ref{gt}) now consists of three independent generation curves.
The generation curve of Source 3 $g_3(t)$ is assumed to be
\begin{equation}
g_3(t)=100.
\end{equation}
$[\alpha(0)]_3=10$.
The generation curve of Source 4 $g_4(t)$ is assumed to be
\begin{equation}
g_4(t)=15+2\cos(40\pi t).
\end{equation}
$[\alpha(0)]_4=15$ and $[\alpha(20)]_4=25$.
The generation curve of Source 5 $g_5(t)$ is assumed to be
\begin{equation}
g_5(t)=5+3\cos(40\pi t)-\sin(40\pi t).
\end{equation}
$[\alpha(0)]_5=20$, $[\alpha(20)]_4=15$ and $[\beta(20)]_5=-25$.
According to the equations in~(\ref{equi}), $\alpha_0'={200\cdot 10+30\cdot 15 +10\cdot 20 \over 240}={265\over 24}$,
$\alpha_{20}'={2\cdot 25+3\cdot 15 \over 5}=19$ and $\beta_{20}'=-25$.

Breakdown of the three payments are summarized in TABLE~\ref{paytbl2}.
We can see that the equivalent price coefficients have conveyed to the demand side the decrease in cost to serve non-dynamic components and $\cos(40\pi t)$ components thanks to a combination of sources compared to the case of single source. They also guarantee the fairness of billing among subscribers. Their total payment is 1445.
The payment to Source 3 is 1000. The payment to Source 4 is 275. The payment to Source 5 is 170. These three payments add up to exactly 1445, indicating that the billing plan is effective.


\begin{table}
\caption{Breakdown of payments based on equivalent price coefficients}
\label{paytbl2}
\centering
\begin{tabular}{|l|c|c|c|}
\hline
\diagbox{Item}{Value}{Load} & Load3 & Load4 & Load5\\
\hline
$a_0$               & 60   & 80  & 100\\
\hline
$a_0$ payment       & 331.25  & 441.67 & 552.08\\
\hline
$a_{20}$            & 15   & 15  & -25\\
\hline
$a_{20}$ payment    & 285  & 285 & -475\\
\hline
$b_{20}$            & 9    & 5   & -15\\
\hline
$b_{20}$ payment    & -225 &-125 & 375\\
\hline
Total               & 391.25  & 601.67 & 452.08\\
\hline
\end{tabular}
\end{table}

\section{Conclusion}\label{clnc}
The cost to match a load curve is more than a matter of how much electric energy is consumed, which uses a number to conclude the curve that belongs to a multidimensional space. It is a matter of the entire fluctuant process of energy consumption. To convey to the demand side the overall effort of power supplier to match with the fluctuant demand load,
we have proposed to analyze load over the entire consumption process, which has been characterized by a square integrable function that belongs an $L_2$ Hilbert space.
We have proposed to determine the payment of electric energy consumption according to the dynamism of a load curve, whereby the load curve is decomposed into a series of
orthonormal basis functions with different periodicity and the dynamism is characterized by amplitudes and polarities of coefficients of the basic functions.
We have proposed to apply the Fourier series to quantify the dynamism and derive the payment for typical connections of a single-bus ideal system.
We have presented examples to demonstrate the effectiveness of the proposed pricing method in conveying the cost of matching load fluctuation and guaranteeing fairness of billing among subscribers and fairness of income distribution among sources.

Rationale of the work, as is worth being stated again,
is to characterize the process of an object per se in space of process to preserve and
exploit the information it contains yielding physical insights, instead of conducting dimension reduction that loses too much significant information.
Dual equivalent spaces of the space of process should be further studied to understand characteristics of the object.
It appeals for more researches on the space of process and its dual spaces to unearth more insights w.r.t. characteristics of an object in continuous operation.

\appendices
\section{Properties of Load Curves}\label{AP1}
We convert the properties of electricity consumption into mathematical expressions.

\emph{Property A.1:}
The addition of load curves satisfies the following:
\begin{enumerate}
\item  For two independent $p_i(t)\in \mathcal{L}$ and $p_j(t) \in \mathcal{L}$, a unique load curve is determined by $p_i(t)+p_j(t)=p_k(t)\in\mathcal{L}$.
\item  $p_i(t)+p_j(t)=p_j(t)+p_i(t)$.
\item  $\left(p_i(t)+p_j(t)\right)+p_l(t)=p_i(t)+\left(p_j(t)+p_l(t)\right)$.
\item There exists an element, denoted by $0$ such that $\forall p(t)\in \mathcal{L}, 0+p(t)=p(t)$.
\item $\forall p(t)\in \mathcal{L}$, there exists an element, denoted by $-p(t)$, such that $p(t)+(-p(t))=0$.
\end{enumerate}

\emph{Property A.2:}
The scalar multiplication of load curves satisfies the following:
\begin{enumerate}
\item  $\forall a \in \mathbb{R}$ and $\forall p(t)\in \mathcal{L}$, $ap(t)\in\mathcal{L}$.
\item  $\forall a_1, a_2 \in \mathbb{R}$ and $\forall p(t)\in \mathcal{L}$, $a_1(a_2p(t))=(a_1a_2)p(t)$.
\item  $(a_1+a_2)p(t)=a_1p(t)+a_2p(t)$.
\item  $a(p_1(t)+p_2(t))=ap_1(t)+ap_2(t)$.
\item  $1\cdot p(t)=p(t)$.
\end{enumerate}

\emph{Property A.3:} $\forall p(t) \in \mathcal{L},$ $$\int_{t_1}^{t_2}p^2(t)dt$$ exists and is finite. It is in accordance with that real load curves are continuous and have finite energy.

Properties 1 to 3 reveal that load curves actually constitute a linear function space as presented in (\ref{set}).

\emph{Property A.4:} The \emph{inner product} on $L_2[t_1,t_2]$ satisfies the following:
\begin{enumerate}
\item  $\Big(p_1(t),p_2(t)\Big)=\Big(p_2(t),p_1(t)\Big)$.
\item  $\Big(p_1(t)+p_2(t),p_3(t)\Big)=\Big(p_1(t),p_3(t)\Big)+\Big(p_2(t),p_3(t)\Big)$.
\item  $\forall a\in \mathbb{R}, \Big(ap_1(t),p_2(t)\Big)=a\Big(p_1(t),p_2(t)\Big)$.
\item  $\Big(p(t),p(t)\Big)\ge 0$. $\Big(p(t),p(t)\Big)=0 \iff p(t)=0$.
\end{enumerate}

\emph{Definition A.1:} The \emph{distance} between two functions $p_1(t)\in L_2[t_1,t_2]$ and $p_2(t) \in L_2[t_1,t_2]$ is defined as:
\begin{equation}
\rho\Big(p_1(t),p_2(t)\Big)=||p_1(t)-p_2(t)||=\sqrt{\int_{t_1}^{t_2}(p_1(t)-p_2(t))^2dt}.
\end{equation}

\emph{Property A.5:} $L_2[t_1,t_2]$ is a Hilbert space, i.e., an inner product space that is also a complete metric space w.r.t. the distance induced by the inner product.
It satisfies the following:
\begin{enumerate}
\item  $L_2[t_1,t_2]$ is an inner product based on \emph{Properties A.1-A.4} and \emph{Definition 1}.
\item  $L_2[t_1,t_2]$ is a complete metric space w.r.t. $\rho\Big(p_1(t),p_2(t)\Big)$.
\item  $L_2[t_1,t_2]$ is a separable space.
\item  $L_2[t_1,t_2]$ is an infinite dimensional space.
\end{enumerate}

\emph{Property A.6:} Load curves $p(t)\in \mathcal{L}$ satisfy the Dirichlet conditions:
\begin{enumerate}
\item $p(t)$ has a finite number of finite discontinuities.
\item $p(t)$ has a finite number of maxima and minima.
\item  $\int_{t_1}^{t_2}|p(t)|dt$ is finite.
\end{enumerate}

%
%
%
%
%



%

\end{document}